\documentclass[10pt,twoside]{article}
\usepackage{amsfonts}
\usepackage{fancyhdr}
\usepackage{titlesec}
\usepackage{cite}
\usepackage{ifthen}
\usepackage{amssymb}
\usepackage{pifont}
\usepackage{stmaryrd}
\usepackage{setspace}
\usepackage{indentfirst}
\usepackage{amsmath,amssymb,amscd,bbm,amsthm,mathrsfs,dsfont}
\input amssym.def

\usepackage{graphicx}

\titleformat{\section}{\centering\large\bfseries}{\S\arabic{section}}{1em}{}
\newboolean{first}
\setboolean{first}{true}

\begin{document}

\setlength\abovedisplayskip{2pt}
\setlength\abovedisplayshortskip{0pt}
\setlength\belowdisplayskip{2pt}
\setlength\belowdisplayshortskip{0pt}

\title{\bf \Large A Novel method to calculate the magnetic field of a Solenoid generated by a surface current element\author{M. Behtouei$^{1}$, B. Spataro$^{1}$, L. Faillace$^{1}$, M. Carillo$^{2}$, M. Comelli$^{3}$,\\ A. Variola$^{1}$,
   and M. Migliorati$^{2, 4}$}\date{}} \maketitle

\noindent {$^{1}$  INFN, Laboratori Nazionali di Frascati, P.O. Box 13,
I-00044 Frascati, Italy}\\
{$^{2}$  {Dipartimento di Scienze di Base e Applicate per l'Ingegneria (SBAI), Sapienza University of Rome, Rome, Italy}}\\
$^{3}${  { Institute of Applied Physics "Nello Carrara", National Research Council, Via Madonna del Piano 10, 50019 Sesto Fiorentino, Italy}}\\
 $^{4}${  {INFN/Roma1, Istituto Nazionale di Fisica Nucleare, Piazzale Aldo Moro, 2, 00185, Rome, Italy }}

\indent  Email: Mostafa.Behtouei@lnf.infn.it

 \footnote{Particle Acceleration, Linear Accelerators, Accelerator applications, Accelerator Subsystems and Technologies, Mathematical Physics }
\begin{center}
\begin{minipage}{135mm}

{\bf \small Abstract}.\hskip 2mm {\small 
The purpose of this paper is to derive the on and off-axes magnetic field of a solenoid with the use of a novel method. We have found a solution for the Biot-Savart law by considering the solenoid with a stationary electric current. The results have been compared to numerical simulations showing a good agreement.
}
\end{minipage}
\end{center}

\thispagestyle{fancyplain} \fancyhead{}

\section{Introduction}
 \label{sec:intro}
The determination of the magnetic field of a solenoid is crucial in physics and engineering. For example, in linear accelerators such as high brilliance photoinjectors and in high gradient accelerating structures \cite{ferrario2007direct,migliorati2009transport,stratakis2010effects,nezhevenko200134,behtouei2020initial}, in presence of the space charge, the effect of beam blow-up has to be thwarted with a solenoid.  Moreover solenoid are used in magnetic correctors and steering devices for beam alignment \cite{brady2013encyclopedia}. 

We extend the results obtained in \cite{behtouei2020novel} by presenting the calculations to determine the magnetic field components of a solenoid with a finite length. Unlike the previous calculation \cite{behtouei2020novel}, in this paper we assume the solenoid is not formed by discrete coils but by a sheet of conductive material so that the current will be distributed on the surface of the solenoid, with a surface current density K. : and comparing the results both with CST particle studio simulation and the analytical results close to the axis of a circular coil.

 \section{Derivation of magnetic induction generated by a surface current element}

Referring to Fig.1, and using cylindrical coordinates, the point P where we want to determine the magnetic field has coordinates (r, $\gamma$, z), while the position P' of the infinitesimal element of the solenoid having surface current density $\vec K$ is given by (r, $\gamma$, $z_0$). The vectors $\vec s'$ and $\vec s$ give the positions of the two points. With this notation, the Biot-Savart law for a solenoid of finite can be written as,

 \begin{equation}\label{mainEQ}
{\bf B}(\mathrm{s})=\frac{\mu_0 }{4\pi}  \int_\Sigma \frac{\vec K \times ({\bf s}-{\bf s'})}{| {\bf s}-{\bf s'}|^{3}} d\Sigma 
\end{equation}

\begin{figure}
 \begin{center}
\includegraphics[width=0.7\linewidth]{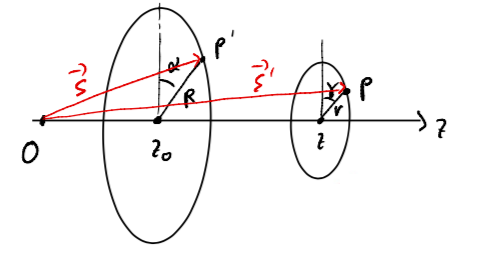}

\caption{ Off-Axis Field Due to a volume current element}
\end{center}
{\small B is the magnetic field at any point in space out of the current loop.

$B_z$:  magnetic field component in the direction of  the coil axis. 

$B_r$:  radial magnetic field component.

$I$ is the current in the wire.

$R$ is the radius of the current loop.

z is the distance, on axis, from the center of the current loop to the field measurement point.

$r$ is the radial distance from the axis of the current loop to the field measurement point

$\alpha$ denotes for the angle of the current element

$\gamma$ stands for the angle of the observer where the magnetic field components are to be calculated}

      \end{figure}

where $\vec K d\Sigma$ is a surface current element  in the direction of current flow and it can be written in cylindrical coordinates:

 \begin{equation}
\vec K=\frac{I}{\ell} \hat \phi.
\end{equation}

 $\ell$ is the length of the solenoid.  By considering $\hat \tau$ the unit vector in the direction of the current of solenoid, the above equation can be written as

 \begin{equation}\label{9}
{\bf B}(\mathrm{s})=\frac{\mu_0 J }{4\pi}  \int_\Sigma \frac{\hat \tau \times ({\bf s}-{\bf s'})}{| {\bf s}-{\bf s'}|^{3}} d\Sigma
\end{equation}

where cross product of  ${\hat \tau}$ and  $({\bf p}-{\bf p'})$ is:

  \begin{multline}\label{15}
  {\hat \tau} \times ({\bf s}-{\bf s'})=
\begin{vmatrix}
    \hat i&\hat j&\hat k \\
  -\sin \alpha&\cos \alpha& 0\\
  \rho\ \cos \gamma-R\ \cos \alpha & \rho\ \sin \gamma-R\ \sin \alpha&z-z_0
\end{vmatrix}\\
=\hat i\ (z-z_0) \ \cos \alpha + \hat j\ (z-z_0)\ \sin \alpha+ \hat k\ (R-\rho\ \cos(\gamma-\alpha)).
\end{multline}

and

\begin{equation}
 |{\bf s}-{\bf s'}|^2=\rho^2+R^2+ (z-z_0)^2-2\rho R \cos (\gamma-\alpha).
\end{equation}

\begin{equation}
d\Sigma=R\ d\alpha \ dz_0
\end{equation}
in the above equation $d\alpha$ denotes the infinitesimal variation of the angle. By some simplification Eq. (\ref{mainEQ}) becomes,

\begin{equation}
{\bf B} (\mathrm{s})=\frac{\mu_0 { K} }{4 \pi} \int_0^{2\pi} \int_{-\ell/2}^{\ell/2} \frac{\hat i\ (z-z_0) \ \cos \alpha + \hat j\ (z-z_0)\ \sin \alpha+ \hat k\ (R-\rho\ \cos(\gamma-\alpha))}{[\rho^2+R^2+ (z-z_0)^2-2\rho R \cos (\gamma-\alpha)]^{3/2}}\ R\ d\alpha \ dz_0
\end{equation}

\begin{equation}
=\frac{\mu_0 { J} }{4 \pi} \int_0^{2\pi} \int_{-\ell/2}^{\ell/2} \frac{(\hat{i} \cos \alpha + \hat{j}\sin \alpha)(z-z_0)+ \vec{k}\ (R-\rho\ \cos(\gamma-\alpha))}{[\rho^2+R^2+ (z-z_0)^2-2\rho R \cos (\gamma-\alpha)]^{3/2}}\ R\ d\alpha \ dz_0
\end{equation}

Writing the components of the magnetic field along the x, y and z coordinates, ${\bf B} (\mathrm{s})=B_x \hat i+B_y \hat j+B_z \hat k$, and then converting them to the cylindrical coordinates we obtain

\begin{equation}
 \mathrm{B}_r={\bf B}\ \cdot \ {\bf r}=  B_x\ \cos \gamma+  B_y\ \sin \gamma
\end{equation}

\begin{equation}
 \mathrm{B}_{\gamma}={\bf B}\ \cdot \ {\bf \tau}=- B_x\ \sin \gamma+  B_y\ \cos \gamma
\end{equation}

where,

\begin{equation}
{\bf \tau}=(-\sin \gamma,\  \cos \gamma,\ 0)
\end{equation}

\begin{equation}
{\bf r}=(\cos \gamma,\  \sin \gamma,\ 0).
\end{equation}

and after some algebraic manipulation we obtain,

\begin{multline}\label{2.5}
 \mathrm{B}_r=\frac{\mu_0 K    }{2 \pi }  \ \int_{- M_{z_\ell}}^{ M_{z_\ell}} (M_z-M_{z_0})  [1+\eta^2 +(M_z-M_{z_0})^2]^{-3/2}  \ dM_{z_0}\\ \int_0^{\pi}   \cos(\psi)[1-\Lambda \ \cos(\psi)]^{-3/2}\  d\psi
\end{multline}

\begin{multline}
 \mathrm{B}_{\gamma}=\frac{\mu_0 K   }{2 \pi } \ \int_{- M_{z_\ell}}^{ M_{z_\ell}} (M_z-M_{z_0})  [1+\eta^2 +(M_z-M_{z_0})^2]^{-3/2}  \ dM_{z_0} \\\int_0^{\pi}   \sin(\psi)[1-\Lambda \ \sin(\psi)]^{-3/2}\  d\psi
\end{multline}

\begin{multline}\label{2.7}
\mathrm{B}_z=\frac{\mu_0 K   R }{2 \pi }  \ \int_{-M_{z_\ell}}^{M_{z_\ell}}  [1+\eta^2+(M_z-M_{z_0})^2]^{-3/2} \ dM_0\\ [\int_0^{\pi}   [1-\Lambda\ cos(\psi)]^{-3/2} d\psi-\eta \int_0^{\pi}    \cos(\psi) [1-\Lambda\ \cos(\psi)]^{-3/2}d\psi]\  
\end{multline}

where $\eta=\rho/R$, $M_z=z/R$ and $\Lambda (R, z, \eta)=2\eta/(1+\eta^2+M_z^2)$. To solve the above integrals we had to deal with two fractional integrals of order 3/2. We solved these integrals by using fractional Cauchy-like integral formula. We cut the created branch line and change the multi-valued operation into the analytic function. The procedure can be found in \cite{behtouei2020novel,behtouei2020application},

\begin{equation}
\int_{0}^{\pi}   [1-\Lambda \ \cos(\psi)]^{-3/2}d\psi
=\pi(1+\Lambda)^{-3/2}\  _2F_1 (\frac{1}{2},\frac{3}{2};1;\frac{2\Lambda}{1+\Lambda})
\end{equation}

\begin{equation}
\int_{0}^{\pi}    \cos(\psi)[1-\Lambda \ \cos(\psi)]^{-3/2}\ d\psi
=-\pi(1+\Lambda)^{-3/2}\  [\ _2F_1 (\frac{1}{2},\frac{3}{2};1;\frac{2\Lambda}{1+\Lambda})- \ _2F_1 (\frac{3}{2},\frac{3}{2};2;\frac{2\Lambda}{1+\Lambda})\ ].
\end{equation}

Where $\ _2F_1 (\frac{1}{2},\frac{3}{2};1;\frac{2\Lambda}{1+\Lambda})$ and $\ _2F_1 (\frac{1}{2},\frac{3}{2};1;\frac{2\Lambda}{1+\Lambda})$ are the ordinary hypergeometric functions having a general form of the kind

\begin{equation}
 \ _pF_q(a_1,..,a_p;b_1,..,b_q;z)=\Sigma_{n=0}^\infty \frac{(a_1)_n ... (b_p)}{(b_1)_n...(b_p)_n}\frac{z^n}{n!}
\end{equation}
where $(a_p)_n, (b_q)_n$ are the rising factorial or Pochhammer symbol with

\begin{equation}\label{2.11}
(a)_0=1
\end{equation}

and

\begin{equation}\label{2.12}
(a)_n=a(a+1)(a+2)...(a+n-1),\ \ \ \ n=1,2,....
\end{equation}

It should be noted that Eq. (\ref{2.5}) is zero due to azimuthal symmetry. By changing of variables Eq.s (\ref{2.5}) and (\ref{2.7}) become,

\begin{equation}
 \mathrm{B}_r=\frac{\mu_0 K  }{2 } \int_{ \sqrt{(1+\eta)^2 + (M_z-M_{z_\ell})^2}}^{ \sqrt{(1+\eta)^2 + (M_z+M_{z_\ell})^2}} \frac{1}{  \beta^{2}}   \   [\  _2F_1 (\frac{1}{2},\frac{3}{2};1;\frac{4 \eta}{\beta^2})- \ _2F_1 (\frac{3}{2},\frac{3}{2};2;\frac{4 \eta}{\beta^2})\ ] \ d\beta
\end{equation}

\begin{equation}\label{2.14}
=\frac{\mu_0 K  }{2  } \ \frac{1}{  \beta}   \   [\  _2\bar F_1 (\frac{3}{2},\frac{1}{2};2;\frac{4 \eta}{\beta^2})- \ _2\bar F_1 (\frac{1}{2},\frac{1}{2};1;\frac{4 \eta}{\beta^2})\ ] |_{ \sqrt{(1+\eta)^2 + (M_z-M_{z_\ell})^2}}^{ \sqrt{(1+\eta)^2 + (M_z+M_{z_\ell})^2}}
\end{equation}

and

\begin{equation}
 \mathrm{B}_z=\frac{-\mu_0 K  }{2 } \int_{ \sqrt{(1+\eta)^2 + (M_z-M_{z_\ell})^2}}^{ \sqrt{(1+\eta)^2 + (M_z+M_{z_\ell})^2}}   \frac{1}{\beta^2(\beta^2-(1+\eta)^2)^{1/2}}  [ \   _2F_1 (\frac{1}{2},\frac{3}{2};1;\frac{4 \eta}{\beta^2})
\end{equation}

\begin{equation}\label{2.15}
+\eta \ (  _2F_1 (\frac{1}{2},\frac{3}{2};1;\frac{4 \eta}{\beta^2})- \ _2F_1 (\frac{3}{2},\frac{3}{2};2;\frac{4 \eta}{\beta^2})\ ) ]  \ d\beta
\end{equation}

with,
\begin{equation}
\beta=[(1+\eta)^2 + (M_z-M_{z_0})^2]^{1/2}
\end{equation}

where $\  _2\bar F_1 (\frac{3}{2},\frac{1}{2};2;\frac{4 \eta}{\beta^2})$ and $ \ _2\bar F_1 (\frac{1}{2},\frac{1}{2};1;\frac{4 \eta}{\beta^2})$ are the regularized  hypergeometric functions having a general form of the kind

\begin{equation}
 \ _p\bar F_q(a_1,..,a_p;b_1,..,b_q;z)=\Sigma_{n=0}^\infty \frac{(a_1)_n ... (b_p)_n}{\Gamma(b_1+n)...\Gamma(b_p+n)_n}\frac{z^n}{n!}
\end{equation}
where $(a_p)_n, (b_q)_n$ are the rising factorial or Pochhammer symbol with $(a)_0$ and $(a)_n$ are the same as Eq.s (\ref{2.11}) and (\ref{2.12}). We could also write the regularized  hypergeometric functions in a simple way $ \ _p\bar F_q(a_1,..,a_p;b_1,..,b_q;z)=(\Gamma(b_1)...\Gamma(b_p))^{-1}  \ _pF_q(a_1,..,a_p;b_1,..,b_q;z) $. To validate Eq.s (\ref{2.14}) and (\ref{2.14}) we would investigate the on axis field components. We observe that the regularized  hypergeometric functions in Eq. (2.14) cancel each other and consequently the radial field component becomes zero. On the other hand, for on axis axial field components we can derive a classic equation which can be found in many physics textbooks and papers \cite{jackson1999classical,bassetti1996analytical,behtouei2020novel}:

\begin{equation}
\mathrm{B}_z=\frac{\mu_0  I R^2 }{2 (R^2+z^2)^{3/2}}.
\end{equation}

The near axis magnetic field components have been expressed by the authors of \cite{bassetti1996analytical} as follows,

\begin{equation}
 \mathrm{B}_r(r, z)=2\Sigma_{p=0}^\infty p\ G_{S2p}(z) r^{2p-1} 
\end{equation}

\begin{equation}
\mathrm{B}_z(r, z)=\Sigma_{p=0}^\infty \ G_{S2p+1}(z) r^{2p} 
\end{equation}

where,

\begin{equation}
G_{S2p}(z)=\frac{\mu_0 I_c}{2R^{2p} Z_L} \Sigma_{k=0}^\infty F_{0,2p,2k+1}\ [f_{2k+1} (Z_L-z)+f_{2k+1} (Z_L+z)]
\end{equation}

\begin{equation}
G_{S2p+1}(z)=\frac{dG_{S2p}(z)}{dz} \ \ and\ \ g_{2k+1} (t)=\frac{df_{2k+1}}{dt} =\frac{(2k+1) R^2}{(R^2+(Z-z)^2)^{3/2}} f_{2k}(t).
\end{equation}

with the following definitions,

\begin{equation}
f_h (t)=(\frac{Z-z}{\sqrt{R^2+(Z-z)^2}})^h, \ \ \ \ (h=0, 1, ... \infty)
\end{equation}

\begin{equation}
F_{0,2p,2k+1}=\frac{(-1)^p}{4^p (p!)^2} (\mathrm{M}^p \mathrm{F}_{00})_{2k+1}
\end{equation}
where M stands for the second derivative of functions $f_{2k+1}$ and $\mathrm{F_{00}}$ is reported in \cite{bassetti1996analytical}. Taking the first two terms of the magnetic field components we obtain:

\begin{equation}\label{66}
 \mathrm{B}_r(r, z)=\frac{\mu_0  J_S }{4} [\frac{3R^2 (Z_L-z)}{(R^2+(Z_L-z)^2)^{5/2}}-\frac{3R^2 (Z_L+z)}{(R^2+(Z_L+z)^2)^{5/2}}] r
\end{equation}

\begin{equation}\label{67}
\mathrm{B}_z(r, z)=\frac{\mu_0  J_S R^2 }{2} \ [ \frac{R^2+(Z_L-z)^2 -3 r^2}{(R^2+(Z_L-z)^2)^{5/2}} - \frac{ R^2+(Z_L+z)^2 -3 r^2}{(R^2+(Z_L+z)^2)^{5/2}} ]
\end{equation}

where the total current of the solenoid being Is, the density function $J_S(Z)$ on its lateral surface is:

\begin{equation}
J_S(Z)=\frac{I_S}{2Z_L} [u(Z+Z_L)-u(Z-Z_L)]
\end{equation}

A solenoid is considered as a set of coils uniformly covering the space between $-Z_L$ and $Z_L$.  Eq. (66) is coincided with Eqs. $\ref{2.14}$ when we consider the first two terms in the regularized expansion of the hypergeometric series $\  _2\bar F_1 (\frac{3}{2},\frac{1}{2};2;\frac{4 \eta}{\beta^2})$ and $\ _2\bar F_1 (\frac{1}{2},\frac{1}{2};1;\frac{4 \eta}{\beta^2})$.

\section{Numerical Results}
\begin{figure}
 \begin{center}
\includegraphics[width=0.5\linewidth]{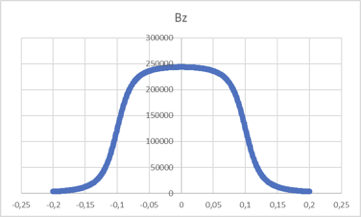}
\caption{ Trend of the Bz component along the axis of the solenoid (the radial component is zero).}
\end{center}

      \end{figure}
WebNIR \cite{webnir} (Web-based tools for assessing occupational exposure to Non-Ionizing Radiation), a portal collecting a series of tools for dissemination and calculation developed as part of a collaboration with other research institutions on exposure to electromagnetic fields, has been developed at the Institute of Applied Physics "Nello Carrara" of the National Research Council (IFAC-CNR). One of these tools, still under development, allows the user to define different types of configurations of conductors (polygonal chains, catenaries, coils, solenoids), define both their geometric and electrical characteristics (current, phase, waveform) and calculate the value of magnetic flux density in correspondence to a set of regularly distributed points in the space (the calculation grid). Two different approaches have been used to perform the calculation: 1) since the field of a segment traversed by current is known, any geometry can be approximated by a polygonal chain consisting of an arbitrarily large number of segments; calculating the field generated by each of them, the resulting field is given by the sum of the individual contributions; 2) for a circular loop, the analytical solution of the field in the space is known \cite{behtouei2020novel,GT} consequently, the field generated by geometries consisting of a set of circular loops (e.g.: coils, solenoids) is determined exactly by considering the elementary contribution of each loop. An "internal validation" of the software results has been carried out by comparing the results obtained by applying, in the case of a solenoid:

- the formulation using the complete elliptic integrals of first and second species \cite{GT}
\begin{figure*}[t]
 \begin{center}
\includegraphics[width=0.4\linewidth]{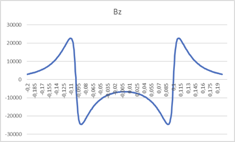}(a)
\includegraphics[width=0.4\linewidth]{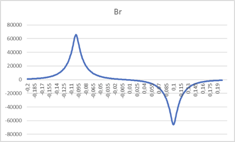}(b)
\includegraphics[width=0.4\linewidth]{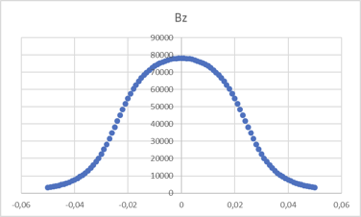}(c)
\includegraphics[width=0.4\linewidth]{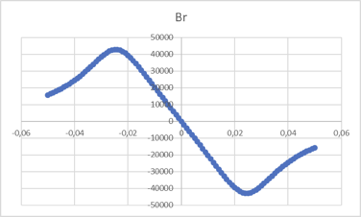}(d)
\includegraphics[width=0.4\linewidth]{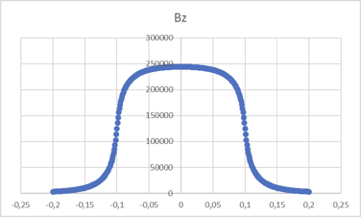}(e)
\includegraphics[width=0.4\linewidth]{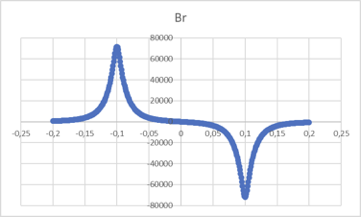}(f)
\caption{ a, b) Trend of Bz and Br 0.5 cm off solenoid. c, d) Trend of Bz and Br orthogonally to the solenoid axis, 1 cm off the inlet. e, f) Trend of Bz and Br 2 cm off axis.}
\end{center}

      \end{figure*}
      
\begin{equation}
\mathrm{B}_r=\frac{\mu_0  I }{2 \pi r \sqrt{z^2+(a+r)^2}   }(\frac{z^2+r^2+a^2}{z^2+(r-a)^2}\ E(k)-K(k))
\end{equation}

\begin{equation}
\mathrm{B}_z=\frac{\mu_0  I }{2 \pi \sqrt{z^2+(a+r)^2}   }(\frac{a^2-z^2-r^2}{z^2+(r-a)^2}\ E(k)+K(k))
\end{equation}

\begin{figure*}[t]
 \begin{center}
\includegraphics[width=0.3\linewidth]{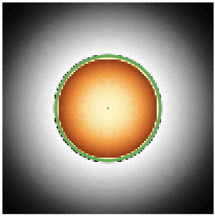}
\includegraphics[width=0.11\linewidth]{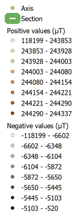}
\caption{ { Distribution of Bz on a cross-section, through the center of the solenoid.}}
\end{center}

 \begin{center}
\includegraphics[width=0.5\linewidth]{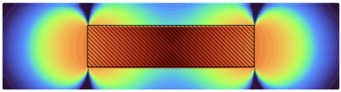}{a}
\includegraphics[width=0.12\linewidth]{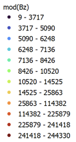}
\includegraphics[width=0.5\linewidth]{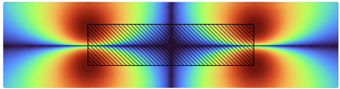}{b}
\includegraphics[width=0.12\linewidth]{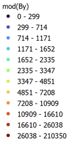}
\caption{ { a) Modulus of Bz on a longitudinal section, through the center of the solenoid. b) Modulus of Br on a longitudinal section, through the center of the solenoid.}
}
\end{center}

      \end{figure*}
 - the formulation that uses the hypergeometric function \cite{behtouei2020novel}
 
 \begin{equation}
\mathrm{B}_r=\frac{\mu_0  I  M_z}{4 \sqrt{2}\pi R  }(\frac{\Lambda}{\eta})^{3/2}\ I_2(\Lambda)
\end{equation}

\begin{equation}
\mathrm{B}_z=\frac{\mu_0  I  }{4\sqrt{2} \pi R } (\frac{\Lambda}{\eta})^{3/2}\ ( I_1(\Lambda)-  \eta\ I_2(\Lambda))
\end{equation}
where,

\begin{equation}
I_1(\Lambda)=\int_{0}^{\pi}   [1-\Lambda \ \cos(\psi)]^{-3/2}d\psi
=\pi(1+\Lambda)^{-3/2}\  _2F_1 (\frac{1}{2},\frac{3}{2};1;\frac{2\Lambda}{1+\Lambda})
\end{equation}

\begin{equation}
\begin{split}
I_2(\Lambda)=\int_{0}^{\pi}    \cos(\psi)[1-\Lambda \ \cos(\psi)]^{-3/2}\ d\psi
\\ =-\pi(1+\Lambda)^{-3/2}\  [\ _2F_1 (\frac{1}{2},\frac{3}{2};1;\frac{2\Lambda}{1+\Lambda})- \ _2F_1 (\frac{3}{2},\frac{3}{2};2;\frac{2\Lambda}{1+\Lambda})\ ]
\end{split}
\end{equation}

We can demonstrate that the magnetic field components obtained by two different methods are identical following the below equations:

 \begin{equation}
 E (\frac{2\Lambda}{\Lambda+1}) =\frac{\pi}{2} \frac{1-\Lambda}{1+\Lambda}\ _2F_1 (1/2,3/2,1,\frac{2\Lambda}{\Lambda+1})
\end{equation}

\begin{equation}
_2F_1 (3/2,3/2,2,\frac{2\Lambda}{\Lambda+1})=\frac{4 E(\frac{2\Lambda}{\Lambda+1})+(\frac{\Lambda-1}{\Lambda+1})K(\frac{2\Lambda}{\Lambda+1})}{\frac{2 \pi \Lambda(1-\Lambda)}{(\Lambda+1)^2})}
\end{equation}

where  K(k) and E(k) are the complete elliptic integrals of the first and  second kinds, respectively.

- the approximate calculation in which the helix which constitutes the solenoid is approximated with a polygonal chain. The number of segments constituting the single loop has been progressively increased in different tests (in detail, the loop was approximated with 20, 200 and 2000 segments).

The result obtained in the first 2 cases is coincident up to the 13th digit: the 2 formulations are equivalent and the discrepancies are due to the approximations implicit in the libraries used. In particular, the language used to perform the calculations is Python 3.9 and the library scipy.special deals with calculating elliptic integrals and hypergeometric functions.

Comparison of these results with those obtained from the approximation with a polygonal chain showed a
progressive convergence as the number of segments with which the helix is approximated increases.

Finally, in the center of a solenoid with the following characteristics: 200 loops, radius=25 mm, distance between loops=1 mm and current intensity=200, the result coincides with that expected from the theoretical formula for the infinite solenoid, given by $B=\frac{\mu_0 N I}{\ell}$ where N indicates the number of loops, I the current, $\ell$ the length of the solenoid and $\mu_0$ is the magnetic permeability in vacuum. With the specified parameters we obtain: $B=8 \times 10^{-2} T\approx 0.25 T$ which is consistent with the numerical results obtained through the software in WebNIR, as shown in the following graphs 2-5.

\end{document}